\documentstyle[11pt]{article}

\newcommand{\beq}{\begin{equation}}
\newcommand{\eeq}{\end{equation}}
\newcommand{\beqa}{\begin{eqnarray}}
\newcommand{\eeqa}{\end{eqnarray}}

\begin{document}

\title{Optical Quantum Random Number Generator}
\author{Andr\'{e} Stefanov, Nicolas Gisin, Olivier Guinnard, Laurent Guinnard, Hugo
Zbinden \\
{\small {\em Group of Applied Physics, University of Geneva, 1211 Geneva,
Switzerland}}}
\date{\today}
\maketitle

\begin{abstract}
A physical random number generator based on the intrinsic randomness of
quantum mechanics is described. The random events are realized by the choice
of single photons between the two outputs of a beamsplitter. We present a
simple device, which minimizes the impact of the photon counters' noise,
dead-time and after pulses.
\end{abstract}

Random numbers are employed today as well for numerical simulations as for
cryptography. Unfortunately computers are not able to generate true random
numbers, as they are deterministic systems. Numerical pseudo-random
generators rely on complexity \cite{Brassard}. Although such pseudo-random
numbers can generally be employed for numerical computation, such as
Monte-Carlo simulations, their use in cryptography, for example to generate
keys, is more critical. The only way to get true random numbers, hence true
security for crypto-systems, is to build a generator based on a random
physical phenomenon [2,3,4]. As quantum theory is intrinsically random, a
quantum process is an ideal base for a physical random number generator.

The randomness of a sequence of numbers can be extensively tested, though
not proven. It is thus of interest to thoroughly understand the behavior of
the random process, so as to gain confidence in its proper random operation.
A statistical process, however, is generally hard to analyze because it
involves a lot of variables. Fortunately, some quantum processes can be well
described with only a few variables, like, for example, the random choice of
a single photon between the two outputs of a beamsplitter [5,6,7]. In this
paper, we present a simple, easy to use and potentially cheap random number
generator based on this quantum process and on the technique of single
photon counting. It fulfills the two major requirements for a physical
random number generator: low correlations between successive outputs and
stability to external perturbations.

The principle of the generator is illustrated in the figure 1. Weak pulses
of a 830 nm LED\ are coupled into a monomode fibre. At the output of the 2
meter long monomode fibre all photons are in the same mode, therefore
indistinguishable, irrespective of any thermal fluctuation of the LED. They
then impinge on two multimode fibres glued together some mm away from the
monomode fiber. Both multimode fibers are coupled to the same photon
counter, one of the multimode fiber introducing a 60 ns delay. By detecting
the time of arrival of the photon one can determine which path it took.
Labeling the short path by '0' and the long by '1' one obtain a sequence of
random bits. The generation rate is of approximately 100kHz, corresponding
to 0.1 photon per pulse as the LED is pulsed at 1MHz. Note that a
Poissonnian photon number distribution with mean number 0.1 is a good
approximation to the ideal single photon delta-distribution. A FPGA circuit
[Xilinx XC 3130] is used to pulse the LED\ and to detect the coincidences.
It features three counters, one for the '0' bits, one for the '1', defined
by two 10 ns large time windows corresponding to the two different arrival
times. The third counter measures the rate of thermal noise thanks to a time
window outside the photon arrival times. The USB port interface to the
computer offers sufficient speed, Plug \& Play support and also the
necessary power supply. The generator fits in a box of small size (68 x 150
x 188 mm).

As photon counter we use a passively quenched Si-APD [EG\&G \ C30902S] in
the Geiger mode \cite{Cova}. The limited efficiency of the detector is not
an issue since the photons which are not detected do not influence the
output of the generator. The thermal noise is not troublesome as it should
be random, but our goal is to avoid such type of statistical random process.
By using a 10 ns coincidence window the contribution of thermal counts is
reduced below 0.5\% even without cooling the detector. More critical are
fast changes of the detector efficiency due to e.g. a ripple on the bias
voltage. Recombining the two optical paths on one detector rather than two
makes the generator much less sensitive to variabilities of the detectors.
Indeed, most of these variations will cancel since they affect in the same
way the '0' and '1' events separated by only 60 ns. However, we have to take
into account the fact that the detector is not in the same state after a
detection as before. Immediately after a detection the bias voltage goes
below breakdown and the efficiency of the detector is zero (dead-time). It
then increases gradually and reaches its original value only after 1 $\mu $%
s. Hence, for pulse frequencies $\geq 1$MHz a two detector scheme would
reveal strong correlations, the probability to detect a '1' after a '0'
being greater than the probability to detect a '0'. In our one detector
scheme this effect is mostly eliminated, but to a small correlation due to
the difference in detection time of 60 ns. This correlation is limited to
the first adjacent bit. It affects only 10\%\ of the bits, as we have
adjacent detection only 10\%\ of the time. The correlation can be further
reduced by decreasing the pulse rate or by electronically rejecting adjacent
detections. We also have to consider the phenomenon of after-pulses: an
increased probability of darkcounts immediately after a detection \cite{Cova}%
. After-pulses decrease with temperature and as we work at room temperature
this effect is not significant.

The raw bits at the output of the generator are not equiprobable, because it
is impossible to achieve a perfect 50/50 coupling between the two optical
paths. But, in order to obtain a 50/50 distribution, one can unbias the bits
by appropriate mathematical procedures. The simplest procedure is that of
von Neumann \cite{Von Neu}, but it's efficiency is limited to 25\%.
Fortunately, there are much more effective procedures, in particular that of
Y. Peres \cite{Peres} which achieves the maximal efficiency given by the
entropy per bit of the sequence. In our prototype the bits are approximately
40/60 distributed and the unbiasing procedure efficiency is greater than
90\%.

In order to check the randomness of the output we applied the
autocorrelation test, which measure the correlation between bits at a
distance $n$: 
\begin{equation}
\Gamma (n)=\frac{1}{N}\sum_{i=0}^{i=N-1}X_{i}\oplus X_{(i+n)~{\rm mod~}N}
\end{equation}
where $\{X_{i}\}_{i=0}^{N-1}$ is a sequence of $N$ bits. Applying this test
on $10^{9}$ raw bits with $1\leq n\leq 2000$, we found no particular
correlation apart from the case $n=1$, which is 5$\sigma $ bellow the mean.
All the other points are normally distributed around the mean value. The
correlation between adjacent bits is very small, on the order of $2\cdot
10^{-4}$. It can be explained by the dead time of the detector, as discuss
above. It disappears when an unbiasing procedure is applied. Other tests,
like frequency, serial and run [11,12], entropy \cite{entropy} and Maurer's
[14,15], did not reveal any deviation from a perfect random source.

In conclusion, we demonstrated a random number generator using a basic
quantum process. Apart from a small correlation between successive bits
which is explained and can be eliminated, the generator behaves like a
perfect random source. As the time delay between the detections
corresponding to a '0' or to a '1' is very small, external perturbations
hardly influence the output of the generator. The prototype is small,
potentially cheap, easy to use \cite{web} and fast enough for cryptographic
applications.

\section*{Acknowledgments}
This work was partly supported by the Esprit project EQCSPOT.

\section*{Figure Caption}
Figure 1: Schematic diagram of the random number generator

\end{document}